\documentclass[11pt,a4paper]{article}
\usepackage{amsmath,amssymb,amsfonts}
\usepackage{graphicx}
\usepackage{url}
\usepackage{epstopdf}
\usepackage{float}
\begin{document}
%


%
%

\title{Temporal Network Analysis of Literary Texts}
{\small
\author{S. D. Prado and S. R. Dahmen\footnote{Corresponding author. E-mail: silvio.dahmen@ufrgs.br}\\
Instituto de F\'{\i}sica da UFRGS\\
91501--970, Porto Alegre, Brazil\\
\\
Ana L.C. Bazzan\\
Instituto de Inform\'atica da UFRGS\\
91501--970, Porto Alegre, Brazil\\
\\
P. Mac Carron\\
Department of Experimental Psychology, University of Oxford\\
Oxford, OX1 3UD United Kingdom\\
\\
R. Kenna\\
Applied Mathematics Research Centre, Coventry University\\
Coventry CV1 5FB  United Kingdom\\
}
}

\maketitle

\begin{abstract}
We study temporal networks of characters in literature focusing on
{\it Alice's Adventures in Wonderland} (1865) by Lewis Carroll and the anonymous {\it La Chanson de Roland} (around 1100). The former, one
of the most influential pieces of nonsense literature ever written, describes the adventures of Alice in a fantasy world with logic plays
interspersed along the narrative. The latter,
a song of heroic deeds,  depicts the Battle of
Roncevaux in 778 A.D. during Charlemagne's campaign on the Iberian Peninsula. We apply methods recently
developed by Taylor {\it et al.}~\cite{Taylor+2015}
to find time-averaged eigenvector centralities, Freeman indices and vitalities
of characters. We show that temporal networks are more appropriate than static ones
for studying {\it stories}, as they
capture features that the time-independent approaches fail to yield.
\end{abstract}

\section{Introduction}

This work was motivated by a simple question: Is there something in the structure of a narrative or in the way characters are
introduced and related to each other, that is crucial to our ability to follow and make sense of a story? Any approach
one may contemplate when trying to `understand' a piece of good literature is of course contingent on the question being asked. 
Finding an answer -- if there is one -- requires much more than a naive analysis that disregards aspects like the
underlying theme, the historical, cultural and sociological contexts at the time of writing and how the readership
relates to these.  Therefore the purpose of the authors is {\it not} to find answers to questions
which can only be tackled with an approach that contemplates all these aspects of literary analysis. Our main objective is
to explore some aspects of the structure of the network of characters of a novel  and see if network theory can be used
as a viable tool in literature. It is known, for instance, that renowned authors construct a
character's {\it persona} as it interacts with other characters and settings~\cite{Truby2007}. Since network theory is the epitome
of a theory of interconnectedness, can it be used to tell us something about how the connections define
one character's importance? Is there a correlation between the mathematical results it yields and our perception,
as readers, of the relevance of this or that character? These are some of the questions we will try to address in
the present work.

Even when one confines oneself to the exclusive use of networks without resorting to other forms of literary analysis~\cite{Eagleton2003}, 
there remain several questions that need to be tackled. The first main challenge is the question of what exactly constitutes
the network in a narrative. Should nodes represent characters or groups of characters, objects or places? What do edges exactly represent?
This question has no simple answer as interactions between characters can be complex and play a fundamental role in the story, as
do objects and locations. In the case of historical texts, narratives depend to a great extent on the narrator and how he/she interprets historical
events~\cite{Gramsch2012}. In spite of having a well-defined theoretical 
framework of `character networks' of Woloch and Moretti~\cite{Woloch2003,Moretti2000,Moretti2005}, choices still have to be made and
there is always more than one possible network for a given narrative~\cite{Moretti2011}.

The  second challenge one faces
when treating networks of actors in a novel or historical narrative is that they are not fixed in time.
Most of the studies on networks have been conducted for systems frozen on a time scale large enough to allow one
to complete disregard any dynamics. All nodes, irrespective of any time ordering regarding the events they depict,
are connected into a single, time-independent network. Following the current terminology, we can call these
{\it aggregate} networks. This may be reasonable in some context
~\cite{Newman2010,Freeman1979,Holme&Saramaki2012,Barrat+2004,Wasserman&Faust1994,Koschtzki+2005,Waerzeggers2014,Padraig&Kenna2012,Padraig&Kenna2013}
but networks involving interpersonal relations and events -- either historical or fictional -- are intrinsically
dynamic~\cite{Gramsch2012,Duering&Stark2011,Holme2015}. Edges may come and go, as well as characters.
Even though one may correctly argue that in a certain sense they are immutable -- history cannot be changed and once
a piece of fiction is written, it remains so -- the stories they depict happen over time, which may typically
span a period of a few days to several decades. If network theory is to have any predictive power, the first question to
come to mind is if  measures which make sense for time-independent networks
will still make sense for dynamical ones and the insights gained from studying the former can still be applied to the
latter~\cite{Holme2015,Holme&Saramaki2013}.

The main goal of the present article is the application of temporal networks to literature and to describe the kinds of questions
one can answer with the methods at hand. To do this we combine the theoretical framework of Woloch~\cite{Woloch2003} and 
Moretti~\cite{Moretti2000,Moretti2005,Moretti2011} for literary studies with techniques developed by Taylor {\it et al.}~\cite{Taylor+2015} for
temporal networks on the mathematical side.  To the best of our knowledge the present paper is the first one where these methods
are applied to literature. We chose two texts as study objects: Lewis Carroll's {\it Alice's Adventures in
Wonderland}~\cite{Carroll1865} of 1865 and the anonymous {\it La Chanson de Roland} from $\it ca.$ 1100 AD~\cite{Roland1100}.
Carroll's {\it Alice} was chosen for several reasons: it has a story with a well defined number of chapters (12), which we took
as being our time sequence. The number of characters is small, which makes it easier to follow an individual character and
see how measures associated with it change over time. It has also another advantage: some of its characters are by now 
household names in many of the 174 languages the book has been translated into. So it is possible to check if some more `popular'
characters, like the Cheshire Cat or the Dodo really stand
the test of a mathematical analysis as to their relevance for the plot. We also looked for a text with a larger set of characters
and which would be as far removed in time and {\it genre} from {\it Alice} as possible: the choice fell on
{\it La Chanson de Roland}, a {\it chanson de geste} (song of heroic deeds)
describing the slaughtering of Roland and his knights. The story is based on the real Battle of Roncevaux of 778,
when the retreating army of Charlemagne was ambushed and the rearguard, to which Roland belonged,
was decimated. The fact that there is an interval of approximately 8 centuries between the two texts
guarantees that any results we obtain is genre-independent and not influenced by the styles characteristics of their time of composition.


This paper is organized as follows: In the first section  we briefly review the theoretical framework of character networks
in literature while addressing the first question: how to choose a network.
This is followed by an exposition of the method propounded in~\cite{Taylor+2015} to deal with time-dependent networks, thus addressing
the second problem: to incorporate the dynamics in the system. We then present the results for the two books in question,
{\it Alice's Adventures in Wonderland} and
{\it La Chanson de Roland}.
To better show the differences when dynamics is considered, we compare these results with
the ones obtained from aggregate, time-independent networks of all characters of each text.
We close the paper with some conclusions and perspectives on the use of network theory to literary texts.

\section{Networks and Literature: Some Background}

In this section we discuss the theoretical aspects of networks in literature while addressing the problem of how to
construct one. The theory of networks of characters, or {\it character networks} as it is called in the specialized literature
was introduced by Moretti~\cite{Moretti2000,Moretti2005,Moretti2011} and
Woloch~\cite{Woloch2003}. An extensive discussion of this theory is presented by Rochat in~\cite{Rochat2014}, who applied
their ideas to study Jean-Jacques Rousseau autobiography {\it Les Confessions} from the viewpoint of networks.
The interested reader should refer to this work for more details. Here we concentrate on
those aspects which are relevant to the present article. According to Rochat, a character network represents
`{\it relations between characters from a text. The relations are based on text proximity, shared scenes/events, quoted speech, etc.}' 
From a literary perspective one starts by defining the concepts of {\it character-space} and {\it character-system}~\cite{Woloch2003}. 
The character-space would be narrative environment of a given character and the character-system the union of all these spaces.
When one picks out one particular network from the character-system, one ends up with character network, which can be
represented as a simple or multiplex, time-dependent or dynamic network. This implies,
of course, making choices as to what relations are important. One key point in the approach is the realization
of the fact that {\it characters in any narrative are defined not by themselves, but as part of a web where each one helps define
the others}. In this sense a story really is a complex system. This is a matter-of-course fact in literary studies but here lies
the link to the application of network theory to narratives:
taking networks as the natural theoretical framework for studying the interdependence between agents that act through space and time, the main task
one faces is how to disregard, in a principled way, aspects of narratives which might not render themselves amenable to a mathematical
treatment. Then comes the question of how data should be collected, by actually reading the book or finding ways of automatically obtaining
information via an adequate software.

Automated network extraction in a context-free way is one option~\cite{Rochat2014,Elson+2010,Sack2013}: one collects characters 
which are mentioned in a page and connect them all into one complete network, even if they do not interact. The procedure
is repeated for the next page and then pages are fused: if edges between two characters appear in both
pages, they are reinforced. New edges are added. This procedure is repeated for all pages, and at the end of the data collecting
one is left with a network which resembles a complete network (everyone connected to everyone else).
This is the approach favoured by~\cite{Rochat2014}. Such context-free method has
its obvious limitations, as for instance placing edges where there might be none. Tolstoy's {\it Anna Karenina} provides
us with a good example of the problems one might run into:
the character Levin and the philosopher Schopenhauer appear repeatedly together, but to the reader it is clear
that Levin speaks {\it of} Schopenhauer but not {\it to} Schopenhauer. Another problem is that this method, in our opinion,
tend to attribute more importance to characters which are peripheral, or keep characters in a plot which at some point of
the narrative disappear. Some ways of avoiding these difficulties or
minimizing them have been discussed in~\cite{Rochat2014}. This approach has its merits, for instance if one is interested
in statistics, when questions regarding not one particular character of a book but literary styles or types of characters are the main focus.
As one does not have the time or resources to read hundreds of books, automated extraction of networks have to be considered a serious
alternative. Another approach is that of Kenna and Mac Carron~\cite{Padraig&Kenna2012,Padraig&Kenna2013}. These authors improved on these ideas by
connecting only those characters who actually meet at some point of the narrative. Of course this requires reading the books and
consequently the number of works that can be tackled is reduced. The character networks of their studies were drawn
from a corpora of myths and sagas as the Irish {\it T\'ain B\'o C\'uailnge}, the Icelandic {\it Sagas}
and the Greek {\it Iliad}. Moreover, being mainly narratives of conflicts, links were also given the attributes {\it friendly} and {\it hostile}.
Without pretending to be exhaustive, these authors calculated a series of measures for the complete, friendly and hostile graphs and
showed that some of the networks in these texts bear a striking resemblance to social networks from the real world, an indication
that the stories they depict could have some elements drawn from real events~\cite{Padraig&Kenna2013}. A more detailed analysis
of the {\it Iliad} beyond that of~\cite{Padraig&Kenna2012} can be found in~\cite{Kydros+2015}. In all the aforementioned studies one does not differentiate
the time in the narrative where the link is made, so there is no dynamics involved.

The importance of dynamics has been discussed by Agarwal {\it et. al}~\cite{Agarwal+2012}. These authors studied {\it Alice's Adventures in Wonderland},
albeit in an online-accessible, abridged version (10 chapters instead of 12). The data extraction was automatic, but checked later by hand.
Their main motivation is in line with multiplex networks, where different
attributes of links changes the relevance of characters according to what is being observed. This can be accomplished by discriminating
between uni/bi-directional and categories: interaction and observation. One character may observe another character
but is not observed by them, as when Alice sees the Rabbit but is not seem by it. The
way they introduced dynamics is by looking at some relevant measures chapter by chapter and comparing them.  Taylor {\it et al.}
devised in~\cite{Taylor+2015} a way of generalizing eigenvector-based measures usually studied in time-independent scenarios to temporal networks.
These authors look at snapshots of the network
at different time layers and treating inter--layer connections (same time) and intra--layer connections (different times)
as being essentially different. For a system with $N$ nodes and $T$ time steps they construct an 
$N\times T$ block diagonal supra-centrality matrix $\mathbb{M}$ where each $N\times N$ diagonal block is the adjacency matrix (or any given function thereof) at a given time step $t=t_1, t_2, \dots t_T$.
A parameter $\varepsilon$ controls how strongly time layers are connected to each other. One may then use all the results from spectral
analysis already known to work for aggregate networks. This ideas were tested with 3 networks:
the exchange of PhD's in mathematics in the United States, the costarring networks of top-billed actors during the Golden Age of Hollywood and citations of
decisions of the American Supreme Court~\cite{Taylor+2015}. The results presented in~\cite{Agarwal+2012} represent a valuable approach towards incorporating
some sort of dynamics into networks as well as in the way they attribute values to edges, but we believe that the methods of Taylor and coworkers
capture the dynamics in a more consistent and rigorous way. We discuss in more detail the differences between these approaches in the next sections.

Since our main goal is to combine character networks with temporal networks, our choices of network building and time reckoning were the following.
As for network extraction, our option follows closely that of Kenna and Mac Carron. As we are interested
in multiplex networks in a {\it temporal} sense and not in a {\it attribute} sense, we consider an edge to exist whenever characters meet
face to face but do not givem them any attribute. This allows us to concentrate on the dynamical aspects of the plot without complicating to
much our analysis. Our choice of time layers was that of chapters, except in the case of {\it La Chanson de Roland}, as we explain below. 
This choice of chapters
seem a reasonable one, as it was made by the author of the narrative and usually contains what one could describe as 'scenes' of a play. For most
stories the scene depicted in a chapter happens {\it after} the scenes of previous chapters, particularly in the case of {\it Alice}.
As for {\it La Chanson de Roland}, one should bear in mind that it was actually recited for an audience and it was thus divided in stanzas or
irregular sizes, called {\it laisses}~\footnote{It is generally believed that this division has to do with the time required for the bard
to rest or play some instrument in between stanzas. For further discussion on the composition of the song, see the introduction to~\cite{Roland1100}.}.
There is a total of 247 to 272 stanzas, depending on the manuscript consulted. As the emphasis
of the story is on the actions rather than on the characters and their introspection, we grouped some stanzas which described a particular scene -- 
a gathering of Charlemagne and his knights or a battle, for example -- into one single chapter. Stanzas describing the scenery were not considered.
With this we ended up with 44 chapters, each one describing a scene.

Going back to the question we discussed in the introduction, we are interested in seeing whether temporal networks can serve as a tool to
select characters according to their relevance to the plot. According to~\cite{Moretti2011,Koschtzki+2005} the narrative importance of a character
can be evaluated by deleting it from the network and comparing to the network before deletion. As one character is removed, so are the links
connected to it. The network becomes more sparse and some minor characters might become completely disconnected from the main narrative.
The quantity which measures this difference is the so-called vitality and it will be discussed in the following section.

\section{Temporal Networks: Some Definitions}
In the theory of static networks, a number of measures have been developed to account for the importance of
nodes~\cite{Wasserman&Faust1994,Newman2010}. Many of these centrality measures
can be expressed in terms of the eigenvector related to the largest eigenvalue of the adjacency matrix~\cite{Taylor+2015,Sola+2013,Benzi&Klymko2015}.
The adjacency matrix  ${\mathbf A}$,
for a network with $N$ nodes is simply defined as the $N\times N$ matrix whose element $ij$ is $1$
if node $i$ and node $j$ are connected or zero otherwise. For a given network, one may simply count
all existing edges, irrespective of when they actually were formed, and define a aggregate adjacency matrix. However, 
if this matrix is to more
realistically represent some network
which evolves in time, so that for instance a given edge between $i$ and $j$ appears at a latter instant of time when, say, nodes $k$ and $l$ are no
longer extant, one has to find ways to represent how the adjacency of a given time changes from time layer to time layer. One way of doing this is
to define a supra-adjacency matrix $\mathbb{M}$ that can be constructed in the following way~\cite{Taylor+2015}: let us assume time is discretized in $T$ steps.
Following~\cite{Taylor+2015} we define $\mathbb{M}$ as the $(NT\times NT)$-matrix 
\begin{equation}
\label{eq:supermatrix}
\mathbb{M} =
  \begin{bmatrix}
    \varepsilon\mathbf{M}^{(1)} & \mathbf{I} & 0 & \cdots  \\
     \mathbf{I} & \varepsilon\mathbf{M}^{(2)} & \mathbf{I} & \ddots  \\
     0 & \mathbf{I}  & \varepsilon\mathbf{M}^{(3)} & \ddots \\
    \vdots & \ddots & \ddots & \ddots 
  \end{bmatrix}
\end{equation}
where each block diagonal $(N\times N)$-matrix $\mathbf{M}^{(t)}$ represents the adjacency matrix $\mathbf{A}(t)$ at a given time
layer $t$ ($t=1,2,\cdots T$). The choice $\mathbf{M}^{(t)}= \mathbf{A}(t)$ is appropriate for our purposes, but $\mathbf{M}^{(t)}$
can be a written as a more general function of $\mathbf{A}(t)$, as for example the hub ($\mathbf{A A^T}$) and authority
($\mathbf{A^T A}$) scores, where $\mathbf{T}$ denotes the transpose matrix~\cite{Taylor+2015}. 
The upper (and lower) diagonal identity matrices are introduced to guarantee that node $i$ at a given
time $t$ is identified with
itself at the next time step $t+1$. The parameter $\varepsilon$ controls how strongly a given node is coupled to itself between neighbouring
time layers. A value of $\varepsilon\rightarrow 0^+$
implies a strong correlation between time layers, meaning order-preserving aggregation. A value of  $\varepsilon\rightarrow\infty$ implies
the decoupling of layers. As shown in~\cite{Taylor+2015}, the limits of small and large values of $\varepsilon$ are well understood.
The centrality trajectories depend on the choices for $\varepsilon$ and there is not a clear 
interpretation for the intermediate regime. However, as pointed out by the authors, the limit $\varepsilon \rightarrow 0^+$ can still give 
valuable information. In this paper, only this latter limit has been explored.

Nodes (characters) are connected by an edge  if they, at some point in the
story, actually meet face to face. If one character talks about another character or thinks of him/her/it, no edge exists.
No value or classification is given to links. Time is measured in chapters. For each chapter (time layer $t$) we calculate
an adjacency matrix $\mathbf{M}^{(t)}$: if, say, characters $i$ and $j$ were connected in previous times but do not contact each other
at time step $t$, the corresponding entry $M_{ij}(t)$ is set to zero. With these partial matrices we build the supra-adjacency matrix $\mathbb{M}$
from which our analysis follows.

In fact the leading eigenvector of $\mathbb{M}$  yields what one calls a joint node-layer centrality,
since it reflects the centrality of both node (character) $i$ and time layer (chapter) $t$.
Since, in general,  these quantities are $\varepsilon$-dependent and not easy to interpret,
the trajectory of node centrality in time is given by the conditional node-layer centrality of the corresponding node-layer $(i,t)$.
The conditional node-layer centrality is the joint node-layer centrality normalized by the MLC of layer t.
Here, MLC stands for {\emph{Marginal Layer Centrality}}, obtained by summing the centrality of all nodes $i$ in a given time layer $t$.

In the limit of  $\varepsilon \rightarrow 0^+$ one expects a slow variation of the conditional node-layer centrality as function of time. 
This is what the authors in~\cite{Taylor+2015} call time-averaged centrality and it ranks nodes so that their centralities are constant in time.
In all  the analyses that follow in this paper, what we call centrality is exactly this time-averaged centrality.
We  should briefly comment on computational complexity here. The supra-centrality matrix given by Eq.(\ref{eq:supermatrix})
whose dominant eigenvector gives the joint node-layer centralities, has size $N T \times N T$ , and that can be problematic
for large networks with many time layers. Conversely, Taylor {\it et al.}  have proved the time-averaged node centralities are given by the
solution of an eigenvalue equation of rank $N$.

We are interested in this paper in the Freeman index of the network. Suppose an unweighted network of $N$
nodes, one particular node $j$ having the highest degree centrality $C_{\text{max}}$ amongst nodes. The degree centrality is the number of
edges connected to a given node. The Freeman index $C_F$ is defined as
\begin{equation}
\label{eq:freemanusual}
 C_F=\frac{\sum_{i=1}^{N}\biggl[ C_{\text{max}}-C_i \biggr]}{(N-1)(N-2)}
\end{equation}
where the sum runs over all nodes with degree centrality $C_i$. The denominador can be explained as follows: the highest centrality a node $j$ can have is in
the so-called star graph, where all nodes are connected to $j$ and to no other node. If this were indeed the case, the centrality of
$j$ would be $C_{\text{max}}=(N-1)$ while $C_i=1$ for $i\ne j$ and the  sum in the numerator of the expression above would be
\begin{eqnarray}
\sum_{i=1}^{N}\biggl[ C_{\text{max}}-C_i \biggr]&=& 0+ (N-1)\times (N-1 -1)\nonumber\\
&=& (N-1)(N-2)
\end{eqnarray}
In our approach, we replace $C_i$ by the eigenvector centrality $c_i$ obtained by diagonalizing $\mathbb{M}$
in the limit $\varepsilon \rightarrow 0^+$ and multiply (\ref{eq:freemanusual}) by a factor $\sqrt{N(N-1)}$ which comes
from the eigenvector normalization. The values of $C_F$ vary between $0$ and $1$, where $1$ corresponds to a star graph while $0$ corresponds to a complete
graph where all nodes have the same number of connections. Geometrically one may interpret the Freeman index as an indicator of how close
a graph is to a star graph. In a literary context this means how `centered' a story is on one particular
character or more `distributed' amongst characters. For example, we might expect a biography to
have a Freeman index closer to $1$, as
the different persons would be mentioned mostly in their connection to the person  being biographed and not in their relation to each
other. Another interesting measure of the relative importance of a given node or its vitality. The vitality
is a concept which measures how some structural properties of a network depend on a node and can be evaluated by deleting it from the network.
In short, by removing a node $i$, another node $j$ emerges as more prominent or is completely disconnect from what is left of the network. To
the best of our knowledge, Moretti was the first to use it in character network analysis. In a case study of Hamlet~\cite{Moretti2011}
he demonstrates the cohesive narrative role of a central actor in that play.
Vitalities can be defined with respect to any real-valued measure $\cal G$ defined over a graph $G$.  The vitality of node $j$ is
given by the difference between $G$ calculated for the whole set of nodes and $G$ for the set of nodes without $j$, namely
\begin{equation}
 V({\cal G}) = {\cal G}(G) - {\cal G}(G\backslash j)
\end{equation}
By deleting a character we mean that all edges attached to it are removed. In our particular case,
we measure the vitality with respect to the eigenvector centralities of nodes.

Recently, Fenu and Higham~\cite{Fenu&Higham2015} have argued that the way the supra-centrality matrix is constructed in eq. (\ref{eq:supermatrix})
can be problematic when one considers directed graphs. This is particularly true in the case of centrality measures based on the concept
of traversals through a network. In their study they consider an alternative formulation where only the identity matrices of the upper
diagonal are present. Since in our case when character A and B meet it is irrelevant who addresses whom, we proceed with the matrix to be given by
(\ref{eq:supermatrix}).

\section{The chapter-by-chapter case}
In what follows, whenever we refer to the books we are studying we will denote them by italics ({\it Alice}, {\it Roland}) whereas
characters will denoted by normal type (Alice, Roland).
Before we present our results for temporal networks
it is interesting to ask whether they are really necessary. Most studies
so far have combined all characters into one single (time-independent) network but according to~\cite{Holme2015}, this can lead
to innacurate results. One could for instance think of a situation where one particular character, highly connected, dies at the
beginning of the story, a fact that an aggregate network does not show at all. One option would be to the go to the opposite extreme,
by taking snapshots of the network at different instants of time and considering them as independent from each other
(in the language of~\cite{Taylor+2015}, this corresponds to the limit of $\epsilon\rightarrow\infty$). 
In order to better illustrate this idea we study {\it Alice}
as being actually 12 networks, one for each chapter. We calculate each character's eigenvector centrality at a given chapter, regardless of what
happened before or happens after that given chapter. This is the approach also used by Agarwal {\it et al.}~\cite{Agarwal+2012} in their study of
{\it Alice}.

\begin{figure}
  \centering 
        \includegraphics[scale=0.9]{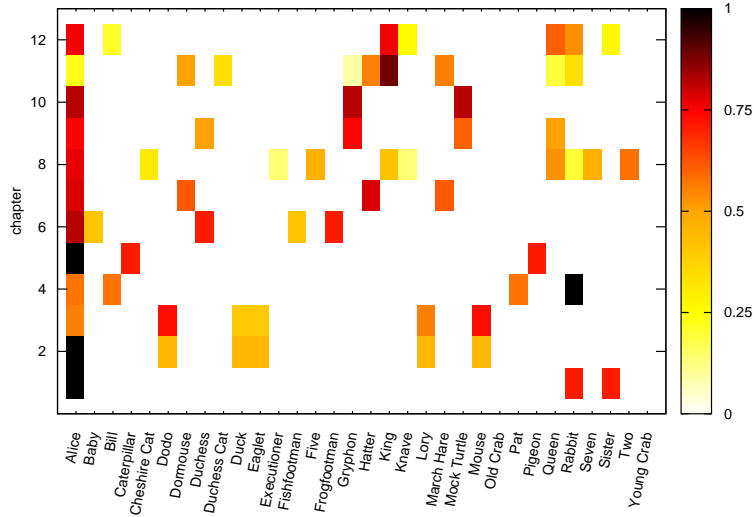}
        \caption{Heat map of characters eigenvector centrality for each of the 12 chapters of {\it Alice's Adventures in Wonderland}.
        Centralities are normalized by chapter.}
        \label{fig:per_chapter}
\end{figure}

The result can be seen in Fig.(\ref{fig:per_chapter}) in the form a temperature map of characters' degrees. In spite of having calculated
different centralities, the graphs of~\cite{Agarwal+2012} are equivalent to this temperature map.
For each chapter, an adjacency matrix is built and diagonalized. The leading eigenvector of this matrix carries the centralities of
each character. The centralities are normalized chapter by chapter.
The result is easy to interpret, as it is equivalent to the impression one gets from actually reading the book, but trying to imagine
that each chapter is a different story. Note that chapter 1 has only three characters: Alice, her sister and the Rabbit, Alice having the highest
degree centrality. In chapter 2 new characters appear while the Rabbit and Alice's sister disappear.
The Rabbit shows up again in chapter 4, while Alice's sister reappears only at the end of the book, when Alice wakes from her dream.
One may clearly see that chapter 8 has the largest number of characters and correspond to the point when Alice arrives at
the  Queen's croquet-ground. By looking at this image one could conclude that the Rabbit appears to be more important than, say,
the Queen, as he appears in more chapters. However one should keep in mind that each centrality is relative to one particular chapter
of the book. This means that in chapter 1, for instance, the Rabbit is being compared to only two other characters, namely Alice and her sister.
So its centrality there cannot be compared to that of the Queen on chapter 8, which has many more characters.
One should also bear in mind that subjectively one identifies the Rabbit in chapter 1 with the same Rabbit in chapter 8 and it
seems natural to define an overall centrality by some sort of weighted or unweighted average. However, from a mathematical point of
view, the characters appearing in each chapter have nothing to do with previous or posterior `reincarnations' of their selves: rigorously they
are completely different actors. What one needs is to find a mathematical way of identifying characters from chapter to chapter, thus
giving the narrative an inner consistency which does not depend our our subjective identification of characters and respects, at the same time,
causality.

\begin{figure}
\centering
        \includegraphics[scale=0.4]{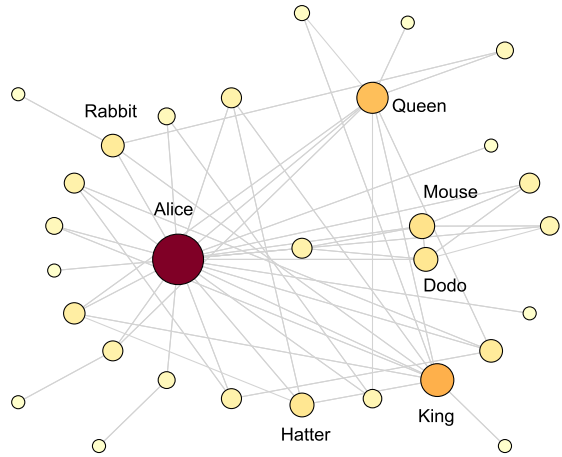}
        \caption{The aggregate network of characters in {\it Alice}. Only some well-known characters are named. Two characters
        are not depicted (the Young and Old Crab) as they do not relate to any other character.}
        \label{fig:alicenet} 
\end{figure}

Previous studies on character networks~\cite{Moretti2011,Rochat2014,Padraig&Kenna2012,Padraig&Kenna2013,Kydros+2015} have used what we call the aggregate network, that is
all connections made along a plot are considered together, irrespective of whether they are, at some point of the narrative, no longer
extant. For the sake of completeness we show what the aggregate
networks of {\it Alice} and {\it Roland} look like in  Figs. (\ref{fig:alicenet}) and (\ref{fig:RolandNet}) respectively. All these studies
however emphasize the importance of incorporating the dynamics of the stories in their studies, something that an aggregate network
does not provide since the causal timeline is lost with the sum of partial results~\cite{Holme2015}. If one expects to get meaningful results,
one must therefore consider the networks in their full time-dependence and identify a given character with itself as the story evolves.
This is exactly what the method of~\cite{Taylor+2015} does, as we show in our case-studies of {\it Alice} and {\it Roland} in what follows.

%
%
%
\begin{figure}
    \centering
        \includegraphics[scale=0.9]{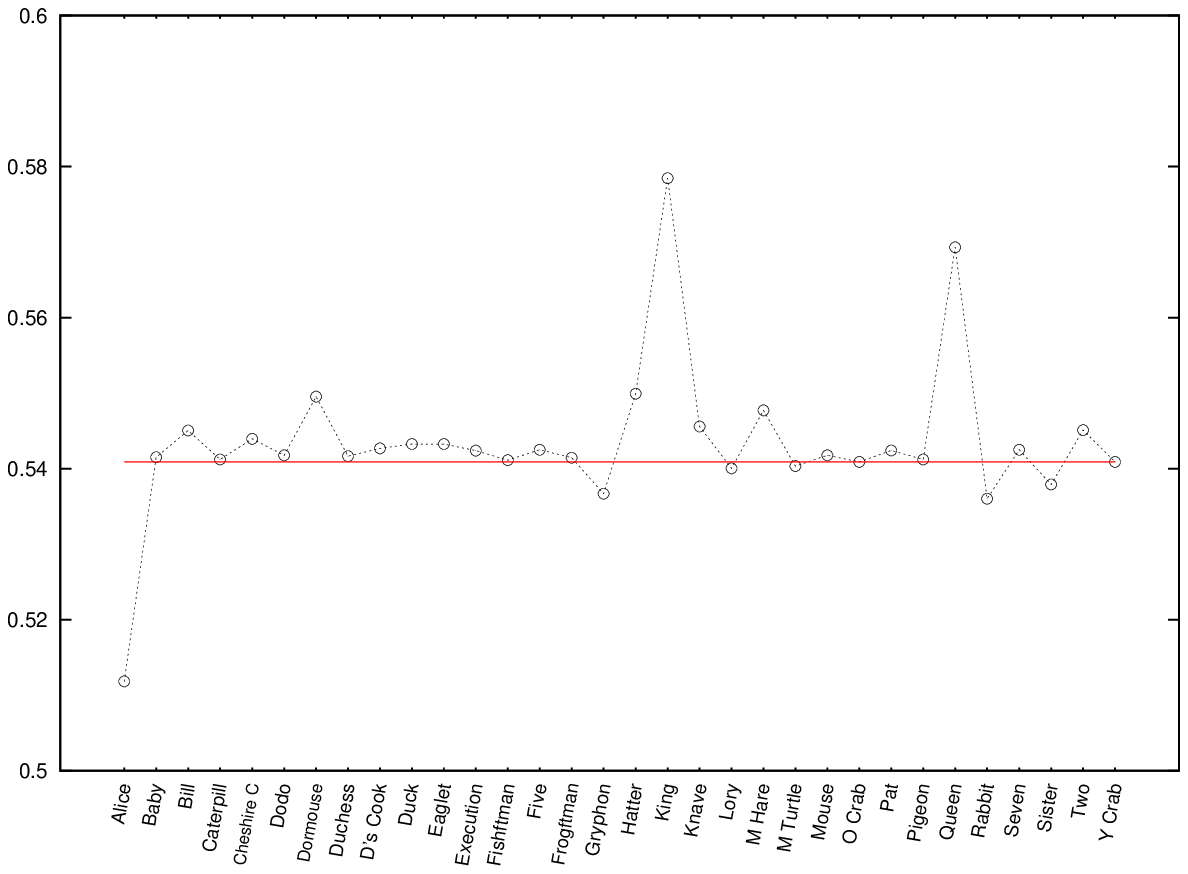}
        \caption{Freeman index calculated for the aggregate case (red line). Dots represent the same quantity when a particular character is
        taken out (named on horizontal axis). The vitality is given by the deviates from the red line.}
        \label{fig:FAE}
\end{figure}
\begin{figure}
\centering
        \includegraphics[scale=0.9]{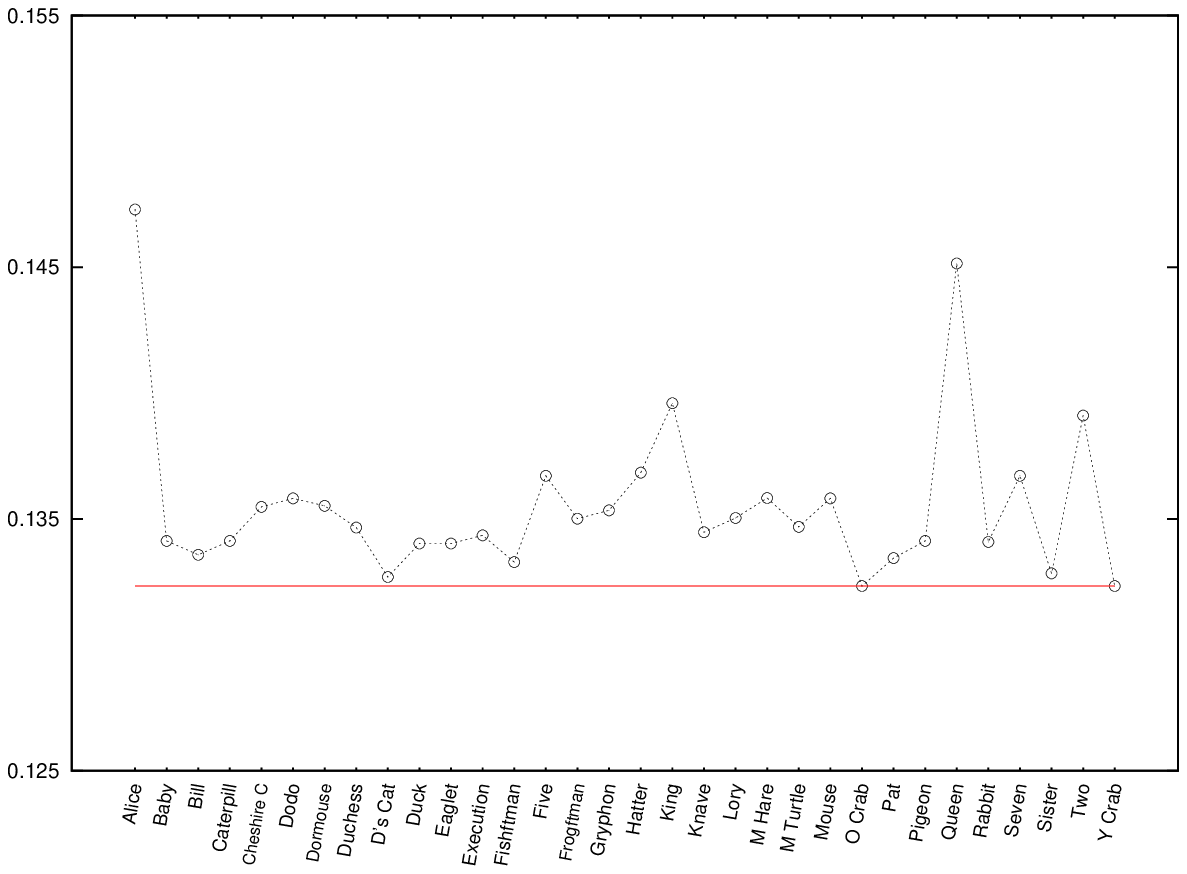}
        \caption{Freeman index calculated for the dynamic case (red line). Dots represent the same quantity when a particular character is
        taken out (named on horizontal axis). The vitality is given by the deviates from the red line.}
        \label{fig:FAP}
\end{figure}

\section{The strong time-coupling limit}

\begin{figure}
    \centering
        \includegraphics[scale=1.0]{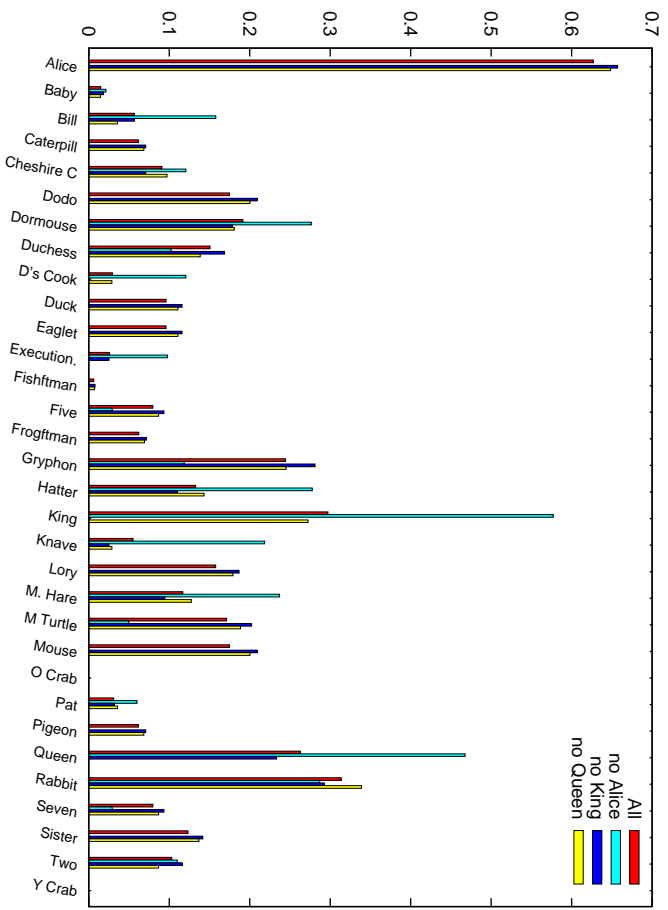}
        \caption{Centrality of characters in {\it Alice} without some of its characters, as indicated by the colors of the
        bars. Aggregate case.}
        \label{fig:CAAE}
\end{figure}
\begin{figure}
\centering
        \includegraphics[scale=1.0]{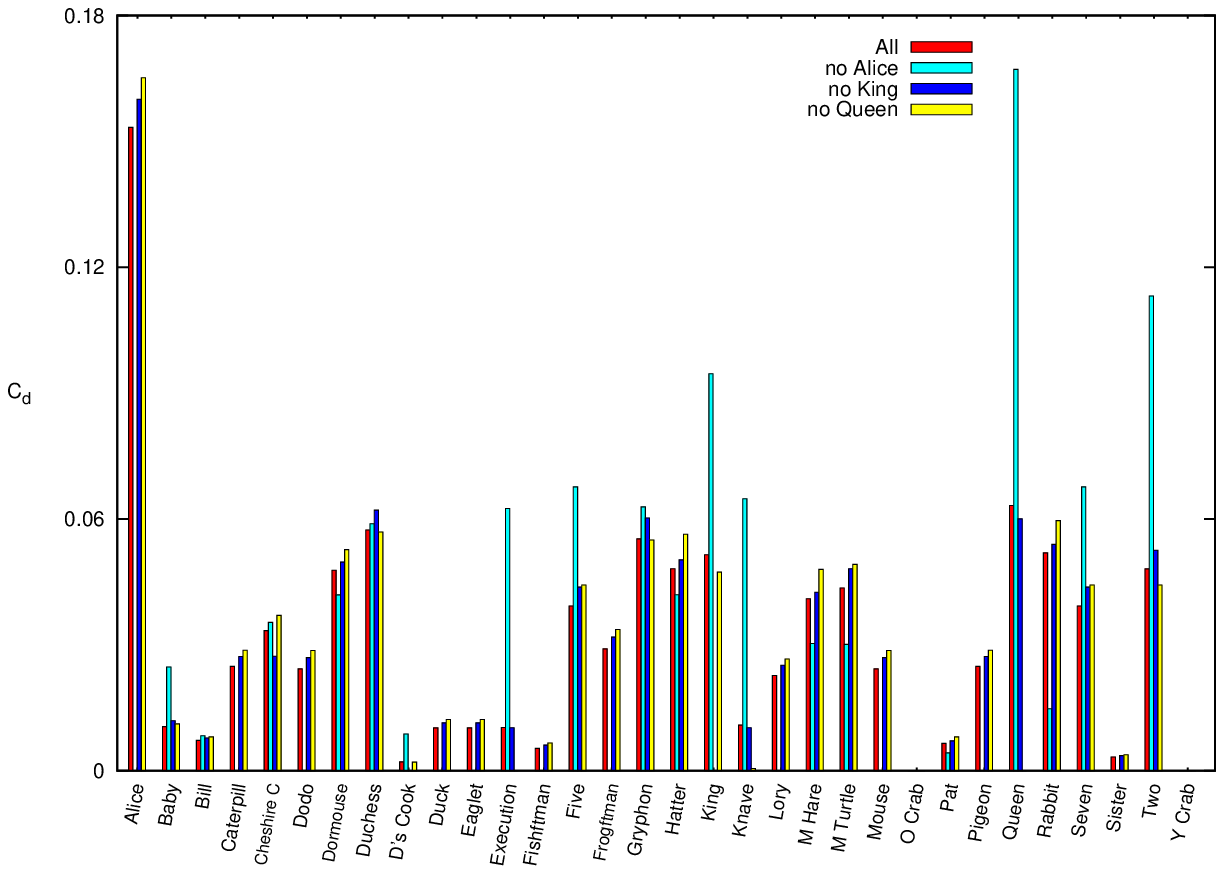}
        \caption{Centrality of characters in {\it Alice} without some of its characters, as indicated by the colors of the
        bars. Dynamical case.}
   \label{fig:AliceCentral}
\end{figure}

We consider now the dynamic case. As mentioned previously the limit of  $\varepsilon \rightarrow 0^+$ 
corresponds to the strong coupling between time layers. In this limit the method presented in~\cite{Taylor+2015} yields a
centrality which is time-independent. To better see the difference
in both cases, we plot in Figs. (\ref{fig:FAE}) and (\ref{fig:FAP}) the Freeman index calculated for {\it Alice} in the (a) aggregate and (b) dynamic
case. The red horizontal line represents the Freeman index of the network. The dots above the name of each character (horizontal
axis) represent the new value of this index when that particular character is taken out of the network. From these two images
it is clear that Alice, the Queen and the King are the characters which most affect the structure of the network. In
the aggregate network, the King and Queen's importance surpasses that of Alice, as the relative change they cause in the Freeman index of
the network is larger. Not only this changes in the dynamic case (Alice becomes more influential) but also the direction of
change is different by the removal of Alice. To understand these differentes we first recall that a smaller Freeman index means that the
network looks more like a complete graph while a higher Freeman index implies that the network is more star-like. If one
compares the aggregate network of Fig. (\ref{fig:alicenet}) with the result depicted in Fig. (\ref{fig:FAE}),
by removing the node Alice the network becomes less star-like, hence the drop in the Freeman index. In the dynamic case, depicted in
Fig.(\ref{fig:FAP}), the change in the index is larger, when compared to the removal of the node Queen or node King. This means that the
node associated with Alice is more relevant as her actions are distributed along the text and not concentrated on a few chapters at the
end, as is the case of the Queen and King. Moreover, for the dynamic case, her removal makes the {\it average} network more star-like as the
remaining networks keep shifting from being more centered on the Queen or the King.

Another way of seeing these results is depicted in Fig. (\ref{fig:AliceCentral}), where we plot what the eigenvector centrality of all characters
would be if some (the most central ones) were to be removed from the network (aggregate and dynamic cases). One can read two things from this graph.
On the one hand, it gives the centrality when everybody is in the story (red bars) and how these change as a certain character is removed
from the story (cyan for a network without Alice, blue without the King and yellow without the Queen). By removing Alice all characters have 
the largest gain compared to their previous values, which implies that Alice has the highest vitality (in both cases, aggregate and dynamic).
However, in the dynamic case, the Queen once again assumes a more prominent role than the King, a fact verified above when considering the
Freeman index of the network.

\begin{figure}[H]
    \centering
        \includegraphics[scale=0.5]{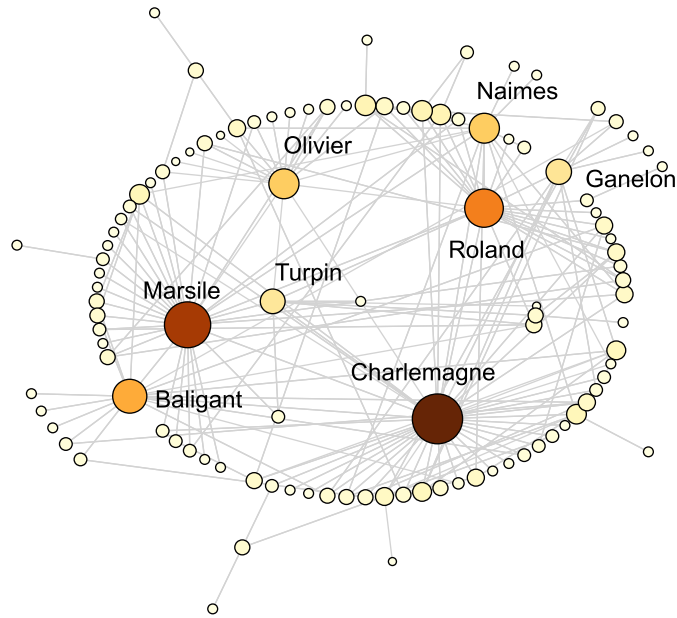}
        \caption{The aggregate network of {\it Roland}. Only better known characters are indicated in the figure.}
        \label{fig:RolandNet}
\end{figure}

One could argue that {\it Alice}'s network is too small and changes are not significant. In order to validate the method in a larger data set we
applied the same ideas to {\it Roland}. In this case the difference between aggregate and dynamic networks become more pronounced, as depicted in
the figures below. For the sake of clarity we also plot the whole aggregate network of Roland in Fig. (\ref{fig:RolandNet}).

For the aggregate network, if one deletes the node associated with Roland the graph becomes
more star-like; one can see in Fig. (\ref{fig:RolandNet}) that the action shifts to well-defined star structure of Charlemagne,
since Roland is not as well connected as Charlemagne. However, in the dynamic case the opposite is actually true: for the narrative,
the absence of the node Roland makes the network more distributed. Roland does not figure as the center of attention,
but subplots become more prominent.
The dynamic network captures the actions and their subplots.

\begin{figure}[H]
\centering
        \includegraphics[scale=0.9]{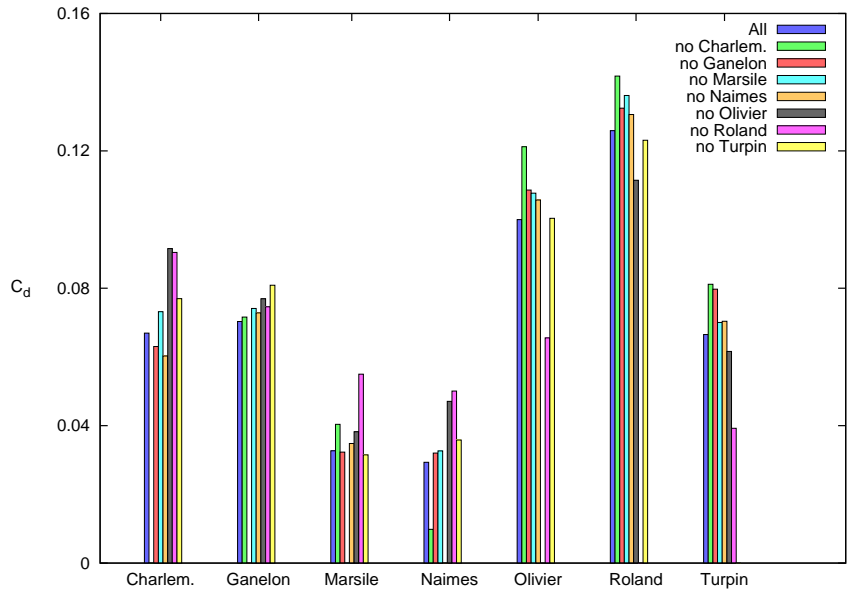}
        \caption{Same figure as above but using~\cite{Taylor+2015} for the dynamic case. Note that in this case Roland becomes the character
        with highest centrality.}
        \label{fig:CARP}
\end{figure}

Fig. (\ref{fig:CARP}) for the eigenvector centrality is more surprising: in comparison to the network of aggregated characters,
which shows Charlemagne as the most connected character ({\it cf.} fig. \ref{fig:RolandNet}), the dynamic case reveals that Roland has
on average a higher eigenvector centrality along the plot. If one thinks in terms of time-independent network this result is
counterintuitive but what it means is that Roland, in spite of having less connections, maintains more of them active during the plot,
thus playing a more prominent role as Charlemagne does.
\begin{figure}[H]
    \centering
        \includegraphics[scale=0.9]{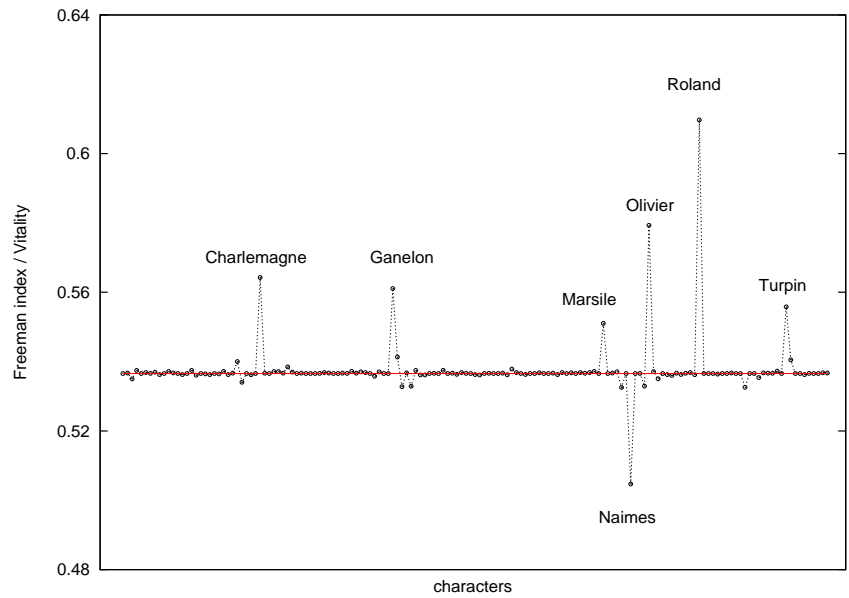}
        \caption{Freeman index of the aggregate network (red line). Dots represent the same quantity when a particular character is
        taken out (named on horizontal axis). The vitality is given by the deviates from the red line.}
        \label{fig:FRE}
\end{figure}
\begin{figure}[H]
\centering
        \includegraphics[scale=0.9]{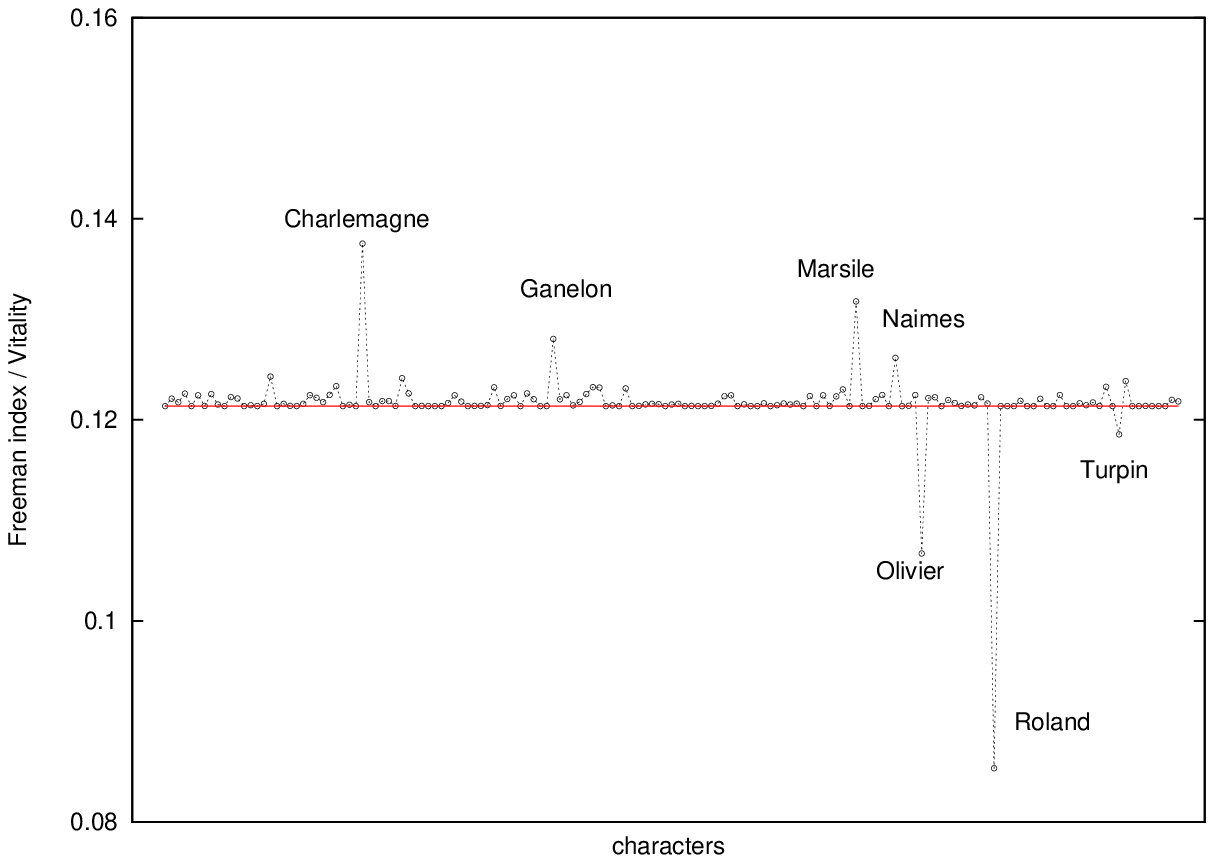}
        \caption{Freeman index of the dynamical network (red line). Dots represent the same quantity when a particular character is
        taken out (named on horizontal axis). The vitality is given by the deviates from the red line.}
        \label{fig:FRP}
\end{figure}

\begin{figure}[H]
    \centering
        \includegraphics[scale=0.9]{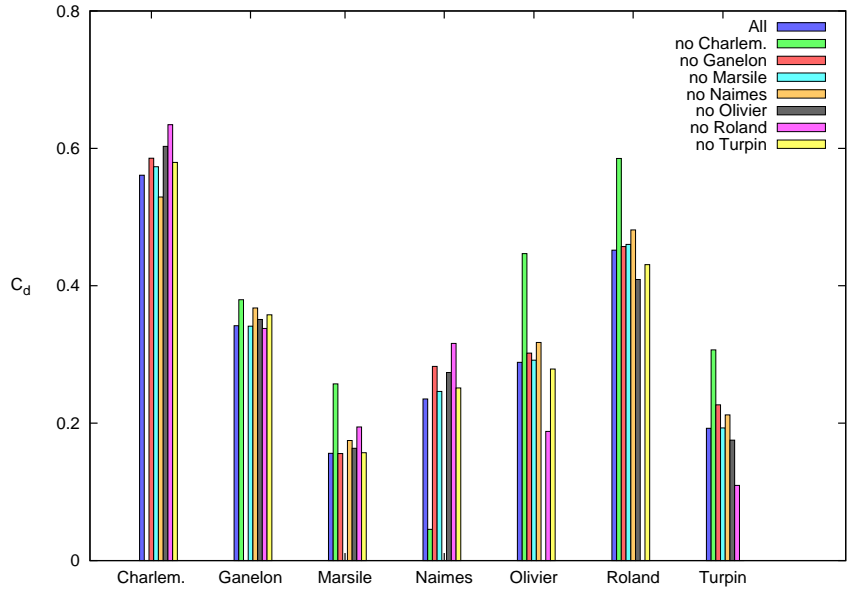}
        \caption{Eigenvector centrality of selected characters of {\it Roland} in the aggregate network. The blue bar represents the
        values obtained when all characters are present. The other bars are the values when some characters are removed from the network
        (color coded according to character).}
        \label{fig:CARE}
\end{figure}

In short, by removing a node $x$, another node $y$ emerges as more prominent for what is left of the network. 
\section{Conclusion}
In this work we applied the methods developed in~\cite{Taylor+2015} for finding eigenvector-based centralities of temporal networks and from these to
determine  the vitality and Freeman index
of characters of two selected literary texts: Lewis Carroll's {\it Alice Adventures in Wonderland}
and the anonymous epic poem {\it La Chanson de Roland}. These two quantities are calculated using the vector-based centrality of a character,
which is given by eigenvector associated with the highest eigenvalue of a supra-centrality matrix $\mathbb{M}$ defined in~\cite{Taylor+2015}. It is
worth emphasizing that this method can be applied  to any centrality that can be expressed as a function of the adjacency matrix of
the network~\cite{Benzi&Klymko2015}.

Our results confirm the utility of this new approach, with results becoming more pronounced
as the network becomes bigger. The application of $\mathbb{M}$  offers new insights into evolving stories, picking out the most important
characters in a way that static networks do not. As for {\it Alice in Wonderland}, the differences between static and dynamic network are
there, but not in such a pronounced way, as the story is rather short (12 chapters) and the number of characters small (32 in total). For
{\it Roland}, however, the dynamic method selects Roland as the most relevant character, in contrast to the static case, where Charlemagne
is the character with highest vitality and Freeman index. Extrapolating these ideas and techniques, we are left to wonder
what kind of interpretations, answers 
or even predictions one would be able to make when applying this approach to model temporal networks in other stories. Any such approach
would of course demand knowning what the `right' questions to ask might be.

As for the amount of subjectivity associated with the construction of a network in literary context, there is no staightforward solution nor
{\it the} right network. As pointed along the text, there are several networks which can be constructed from a set of nodes.
The connections are contigent on the question one is trying to answer. This does not make network theory {\it a priori} an invalid tool
for the analysis of complex relations between actors in a given setting. On the contrary, comparison between what network theory
predicts and reality should be used as a criterium for the validity of this approach, as we hoped to have shown in the results we presented.

Finally, there exists an enormous amount of texts which could be analyzed. It is of course an impossible task to read all of them, but if
one is able to devise a way of systematically gathering information from classics, network theory, in particular temporal networks, may provide
a kind of network-theoretical signature to classify authors, genres and epochs.   

\section*{Acknowledgments}
A.L.C.B, S.R.D. and S.D.P. were supported by IRSES Grant Project PIRSES-GA-2011-295302. The hospitality of the AMRC in Coventry is
gratefully acknowledged. PMC was supported by a European Research Council Advanced grant to R.I.M. Dunbar.


\end{document}